\documentclass[graybox, envcountchap]{svmult}

\newcommand{\Xe}{{X_{\rm e}}}
\newcommand{\LCDM}{{$\Lambda{\rm CDM}$}}
\newcommand{\aEM}{\alpha_{\rm EM}}
\newcommand{\WMAP}{\emph{WMAP}\xspace}
\newcommand{\Planck}{\emph{Planck}\xspace}

\newcommand{\Mpc}{{\rm Mpc}}

\newcommand{\expf}[1]{{{\rm e}^{#1}}}

\newcommand{\nS}{n_{\rm S}}

\newcommand{\id}{{\,\rm d}}

\newcommand{\beq}{\begin{equation}}   %

\newcommand{\eeq}{\end{equation}}   %

\newcommand{\beqa}{\begin{eqnarray}}   %

\newcommand{\eeqa}{\end{eqnarray}}   %

\newcommand{\beal}{\begin{align}}
\newcommand{\enal}{\end{align}}

\newcommand{\bspl}{\begin{split}}

\newcommand{\espl}{\end{split}}

\newcommand{\bsub}{\begin{subequations}}

\newcommand{\esub}{\end{subequations}}

\newcommand{\bmulti}{\begin{multline}}   %

\newcommand{\beqm}{\begin{mathletters}}   %

\newcommand{\eeqm}{\end{mathletters}}   %

\newcommand{\me}{m_{\rm e}}

\newcommand{\Ne}{N_{\rm e}}

\newcommand{\Te}{T_{\rm e}}

\newcommand{\Tg}{T_{\gamma}}

\newcommand{\sigT}{\sigma_{\rm T}}

\usepackage{xspace}
\usepackage{mathptmx}        
\usepackage{amsmath}
\usepackage{amssymb}
\usepackage{color}
\usepackage{helvet}          
\usepackage{courier}         
\usepackage{dirtree}

\usepackage{makeidx}        
\usepackage{graphicx}        
\usepackage{subfig}

\usepackage{multicol}        
\usepackage[bottom]{footmisc}

\usepackage{hyperref}        
\hypersetup{colorlinks=true,urlcolor=blue}

\usepackage[misc]{ifsym}

\makeindex             

\newcommand{\dalpha}{\Delta\alpha/\alpha}
\newcommand{\dme}{\Delta m_{\rm e}/m_{\rm e}}
\newcommand{\aEMs}{\alpha_{\rm EM, 0}}
\newcommand{\mes}{m_{\rm e, 0}}

\newcommand{\EDEC}{\cite{Poulin2023Chapter, Raveri2023Chapter}}
\newcommand{\PMFC}{\cite{Jedamzik2023Chapter}}

\begin{document}


\title{Varying fundamental constants meet Hubble}
\titlerunning{Varying fundamental constants}
\author{Jens Chluba and Luke Hart}
\institute{Jens Chluba (\Letter) \at Jodrell Bank Centre for Astrophysics, Alan Turing Building, University of Manchester, Manchester M13 9PL, United Kingdom
\\
\email{jens.chluba@manchester.ac.uk}
\and Luke Hart (\Letter) \at Jodrell Bank Centre for Astrophysics, Alan Turing Building, University of Manchester, Manchester, M13 9PL, United Kingdom
\\
TNEI Services Ltd., Bainbridge House, 88 London Road, Manchester, M1 2PW, United Kingdom
\\
\email{luke.hart@tneigroup.com}
}

\maketitle

\abstract{Fundamental physical constants need not be constant, neither spatially nor temporally. -- This seeming simple statement has profound implications for a wide range of physical processes and interactions, and can be probed through a number of observations. 
In this chapter, we highlight how CMB measurements can constrain variations of the fine-structure constant and the electron rest mass during the cosmological recombination era.
The sensitivity of the CMB anisotropies to these constants arises because they directly affect the cosmic ionization history and Thomson scattering rate, with a number of subtle atomic physics effects coming together. 
Recent studies have revealed that variations of the electron rest mass can indeed alleviate the Hubble tension, as we explain here. Future opportunities through measurements of the cosmological recombination radiation are briefly mentioned, highlighting how these could provide an exciting avenue towards uncovering the physical origin of the Hubble tension experimentally.}


\newpage
\section{Why varying fundamental constants?}
\label{sec:intro}
The laws of nature depend on fundamental constants (FCs) such as Newton's constant, $G$, the speed of light, $c$, and elementary charge of an electron, $e$, to name just a few. The values of these constants have been determined experimentally, but generally should emerge directly from the underlying theory. As such, there is no reason to assume that the values of the FCs determined locally simply translate to other parts of the cosmos or to other eras in cosmic history \cite{Uzan2003,Uzan2011,Martins2017}. Studies of fundamental constants and their possible temporal and spatial variations are thus of utmost importance, and could provide a glimpse at physics beyond the standard model, possibly shedding light on the presence of additional scalar fields and their couplings to the standard sector.

Of the many fundamental constants, the fine-structure constant, $\aEM$, and electron rest mass, $\me$, are the most interesting to CMB studies \cite{Kaplinghat1999, Avelino2000, Battye2001, Avelino2001, Rocha2004, Martins2004, Scoccola2009, Menegoni2012}. This is because these constants play crucial roles in the way photons and baryons interact. Most importantly, they affect important atomic transition rates, which in turn control the cosmological recombination process and hence the Thomson visibility function (defining the last scattering surface) that is so crucial to the formation of the CMB temperature and polarization anisotropies \cite{Sunyaev1970, Peebles1970, Hu1995CMBanalytic, Hu1995, Lewis2006}. 

Using {\it Planck} 2013 data, the values of $\aEM$ and $\me$ around recombination were proven to coincide with those obtained in the lab to within $\simeq 0.4\%$ for $\aEM$ and $\simeq 1\%-6\%$ for $\me$ \cite{Planck2015var_alp}. These limits are $\simeq 2-3$ orders of magnitude weaker than constraints obtained from other `local' measurements \cite{Bize2003,Rosenband2008, Bonifacio2014, Kotus2017}; however, the CMB places limits during very different phases in the history of the Universe, centered around the time of last scattering some $380,000$ years after the Big Bang, thereby complementing these low-redshift measurements. In addition, CMB measurements can be used to probe spatial variations of the FCs at cosmological distances \cite{Smith2019}, opening yet another avenue for exploration.

With the \Planck 2013 results in mind, no significant surprises were expected from the analysis of improved CMB data of the {\it Planck} 2015 and 2018 releases. However, it turns out that when considering models with varying $\me$, the geometric degeneracy becomes significant and can accommodate shifts in the value of the Hubble parameter when multiple probes are combined \cite{Hart2020Hubble}. The same geometric freedom is not encountered when varying $\aEM$ due to the modified dependence of the visibility function on this parameter. This finding has spurred an increased interest in studying VFCs in this context, with scenarios that allow for varying $\me$ \cite{Sekiguchi2021PhRvD, Lee2023VFC, Tohfa2023, Hoshiya2023} ranking high in model comparisons \cite{NilsH0Olympics}. Since VFCs can be caused by the presence of scalar fields \cite{Bekenstein1982, Martins2015}, a natural question is whether the same scalar fields could also be causing effects relating to EDE (see \EDEC), indicating a 'two sides of the same coin' interplay. It will therefore be extremely important to ask how different measurements can be combined to shed light on the physical origin of the Hubble tension.

In this Chapter we explain how $\aEM$ and $\me$ enter the cosmological recombination problem and calculation of the CMB power spectra. We then briefly recap some of the constraints, highlighting important findings, before moving on to a discussion of their role in the Hubble tension. Modifications to the recombination process also affect the cosmological recombination radiation (CRR) \cite{Sunyaev2009, Chluba2016CosmoSpec}, implying that direct insight into the underlying physics could be gained by future measurements of CMB spectral distortions \cite{Chluba2021Voyage}. We highlight how this avenue may even allow distinguishing models of EDE, PMFs and VFCs and possibly identify modifications to the recombination history as the main cause of the tension \cite{Hart2023VFCCRR, Lucca2023}.  

\section{Effects of VFCs on the recombination process}
\label{sec:recombination_VFC}
The ionization history of the Universe is one of the crucial theoretical ingredients in the computations of the CMB temperature and polarization anisotropies \cite{Sunyaev1970, Peebles1970, Hu1995}. The first computations of this transition from the fully-ionized plasma to a neutral medium were carried out in the late 60s, recognizing the important role of Lyman-$\alpha$ transport and the 2s-1s two-photon decay of Hydrogen \cite{Zeldovich68, Peebles68}. These early calculations reached a precision of $\simeq 20\%-30\%$ in the free electron fraction, $\Xe=N_{\rm e}/N_{\rm H}$, around the maximum of the Thomson scattering visibility function at redshift $z\simeq 1100$. Here $N_{\rm e}$ is the free electron number density and $N_{\rm H}$ denotes the total number density of hydrogen nuclei in the Universe. However, with the advent of precision CMB data from \WMAP and \Planck it became important to improve the modeling of the cosmological recombination process \cite{Hu1995, Chluba2006, Lewis2006}. Initially, this led to the development of {\tt Recfast} \cite{SeagerRecfast1999}, which reached $\simeq 1\%-3\%$ precision for Hydrogen and neutral Helium recombination. However, for the analysis of data from \Planck, this precision was still insufficient, and many detailed atomic physics and radiative transfer effects had to be accounted for \cite{Fendt2009, Sunyaev2009, Wong2008, RubinoMartin2010, Glover2014}, leading to changes at the level of several standard deviations in particular for the value of the spectral index of scalar perturbations, $\nS$ \cite{RubinoMartin2010, Shaw2011}. 
This necessitated the development of the highly-flexible and accurate recombination codes {\tt CosmoRec} \cite{Chluba2010b} and {\tt HyRec}~\cite{Yacine2010c}, which ensured that for the analysis of \Planck none of the standard parameters were biased at a significant level \cite{Planck2015params, Planck2018params}.

This short recap highlights the crucial role of the recombination history in the computations of the CMB anisotropies using standard Boltzmann solvers such as {\tt CAMB} \cite{CAMB} and {\tt CLASS} \cite{CLASSCODE}, and conversely, the impressive precision and sensitivity of the current measurements to subtle modifications in the ionization history. The statements assume standard physics during the recombination era at $z\simeq 10^2-10^4$. In particular, it is assumed that the atomic transition rates for Hydrogen and Helium are the same as those inferred in the lab. 

From standard textbook atomic physics, it is well-understood how 
$\aEM$ and $\me$ affect the energy levels and transition probabilities of Hydrogen and Helium \cite{Bethe1957, DrakeBook2006}.
It is immediately clear that varying $\aEM$ and $\me$ inevitably create changes to the cosmological ionization history and hence CMB observables. Most importantly, the energy levels of hydrogen and helium depend on these constants as $E_i\propto\aEM^2\me$, which directly affects the recombination redshift. In addition, the atomic bound-bound transition rates and photoionization/recombination rates are altered when varying $\aEM$ and $\me$. Lastly, the interactions of photons and electrons through Compton and resonance scattering modify the radiative transfer physics, which control the dynamics of recombination \cite{Chluba2007b, Kholupenko2007, Jose2008, Hirata2008, Chluba2008b, Yacine2010b, Chluba2012HeRec}. 

In an effective three-level atom approach \cite{Zeldovich68, Peebles68, Seager2000}, the individual dependencies can be summarized as \cite{Kaplinghat1999, Scoccola2009, Planck2015var_alp, Hart2017}
\begin{align}
\label{eq:recfast_scaling}
\begin{split}
\sigma_{\rm T} \propto \aEM^2 \me^{-2} 
\qquad A_{2\gamma} &\propto \aEM^8 \me 
\qquad P_{\rm S} A_{1\gamma} \propto \aEM^{6}\me^{3} 
\\
\alpha_{\rm rec} \propto \aEM^2 \me^{-2} 
\qquad \beta_{\rm phot} &\propto \aEM^5 \me 
\qquad T_{\rm eff} \propto \aEM^{-2}\me^{-1}.
\end{split}
\end{align} 
Here, $\sigma_{\rm T}$ denotes the Thomson scattering cross section; $A_{2\gamma}$ is the two-photon decay rate of the second shell; $\alpha_{\rm rec}$ and $\beta_{\rm phot}$ are the effective recombination and photoionization rates, respectively; $T_{\rm eff}$ is the effective temperature at which $\alpha_{\rm rec}$ and $\beta_{\rm phot}$ need to be evaluated (see explanation below); $P_{\rm S}A_{1\gamma}$ denotes the effective dipole transition rate for the main resonances (e.g., Lyman-$\alpha$), which is reduced by the Sobolev escape probability, $P_{\rm S}\leq 1$ \cite{Sobolev1960, Seager2000} with respect to the vacuum rate, $A_{1\gamma}$.
For a more detailed account of how the transition rates depend on the fundamental constants we refer to {\tt CosmoSpec} \cite{Chluba2016CosmoSpec} and the manual of {\tt HyRec} \cite{Yacine2010c}. 

The scalings of $\sigma_{\rm T}$, $A_{2\gamma}$ and $P_{\rm S}A_{1\gamma}$ directly follow from their explicit dependence on $\aEM$ and $\me$. For $\alpha_{\rm rec}$ and $\beta_{\rm phot}$, only the renormalisations of the transition rates is reflected, again stemming from their explicit dependencies on $\aEM$ and $\me$ \cite{Karzas1961}. However, these rates also depend on the ratio of the electron and photon temperatures to the ionization threshold. This leads to an additional dependence on $\aEM$ and $\me$, which can be captured by evaluating these rates at a rescaled temperature, with a scaling indicated through $T_{\rm eff}$.
Overall, this leads to the effective dependence $\alpha_{\rm rec}\propto \aEM^{3.44} \, \me^{-1.28}$ around hydrogen recombination \cite{Chluba2016CosmoSpec}. The required photoionization rate, $\beta_{\rm phot}$, is obtained using the detailed balance relation. Slightly different overall scalings for $\alpha_{\rm rec}$ and $\beta_{\rm phot}$ were used in \cite{Planck2015var_alp}, but the associated effect on the recombination history were found to be sub-dominant and limited to $z\lesssim800$ \cite{Hart2017}.

For neutral helium, non-hydrogenic effects (e.g., fine-structure transitions, singlet-triplet couplings) become relevant \cite{DrakeBook2006, Drake2007}. However, the corrections should be sub-dominant and are usually neglected. A detailed discussion of changes to the escape probabilities during helium recombination can be found in \cite{Hart2023VFCCRR}. In \cite{Hart2017}, the changes to helium and hydrogen recombination were furthermore treated separately.

\subsection{Ionization history modifications due to variation of $\aEM$}
\label{sec:Xe_aEM}
\begin{figure}[t]
	\centering
 \includegraphics[width=0.85\columnwidth]{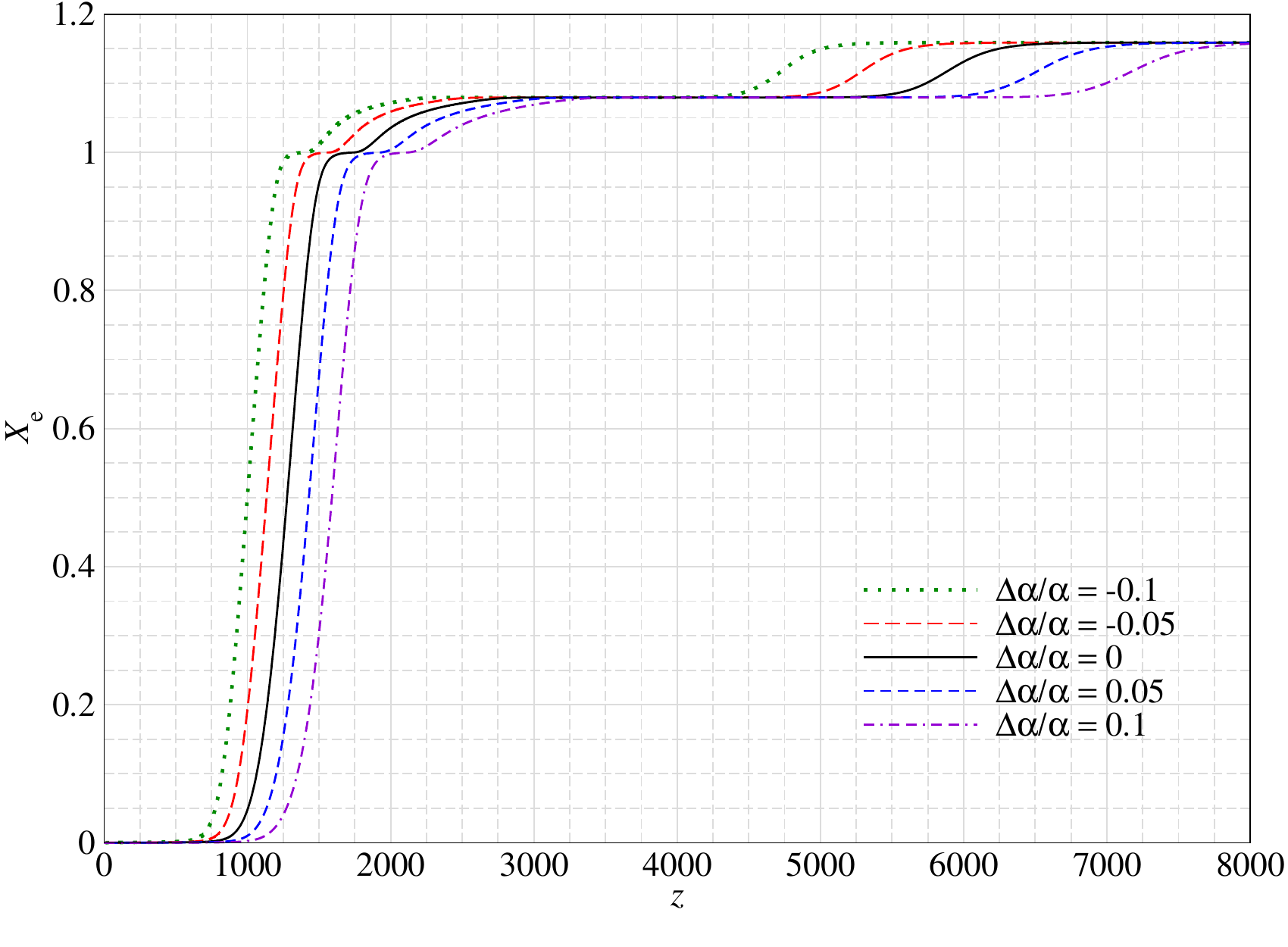}
    \caption{Ionization histories for different values of $\aEM$. The dominant effect is caused by modifications of the ionization threshold, which implies that for increased $\aEM$ recombination finishes earlier. The curves were computed using the {\tt Recfast++} module of {\tt CosmoRec} \cite{Chluba2010b}. The figure was taken from \cite{Hart2017}.}
    \label{fig:alpha}
\end{figure}

Given the above ingredients, one can now answer the question about how various effects propagate to the ionization history. A detailed study that also directly demonstrated the validity of simpler three-level approximations against {\tt CosmoRec} was carried out in \cite{Hart2017}.
The overall effect of varying $\aEM$ is illustrated in Fig.~\ref{fig:alpha}, assuming a constant variation parametrized as $\aEM=\aEMs(1+\dalpha)$, where $\aEMs=1/137$ denotes the standard value. 
Increasing the fine structure constant shifts the moment of recombination toward higher redshifts. This agrees with the results found earlier in \cite{Kaplinghat1999, Battye2001, Rocha2004} and can intuitively be understood in the following manner: $\dalpha>0$ increases the transition energies between different atomic levels and the continuum. This increases the energy threshold at which recombination occurs, hence increasing the recombination redshift, an effect that is basically captured by an effective temperature rescaling in the evaluation of the photoionization and recombination rates (see below). 

\begin{figure}
	\centering
 \includegraphics[width=0.85\columnwidth]{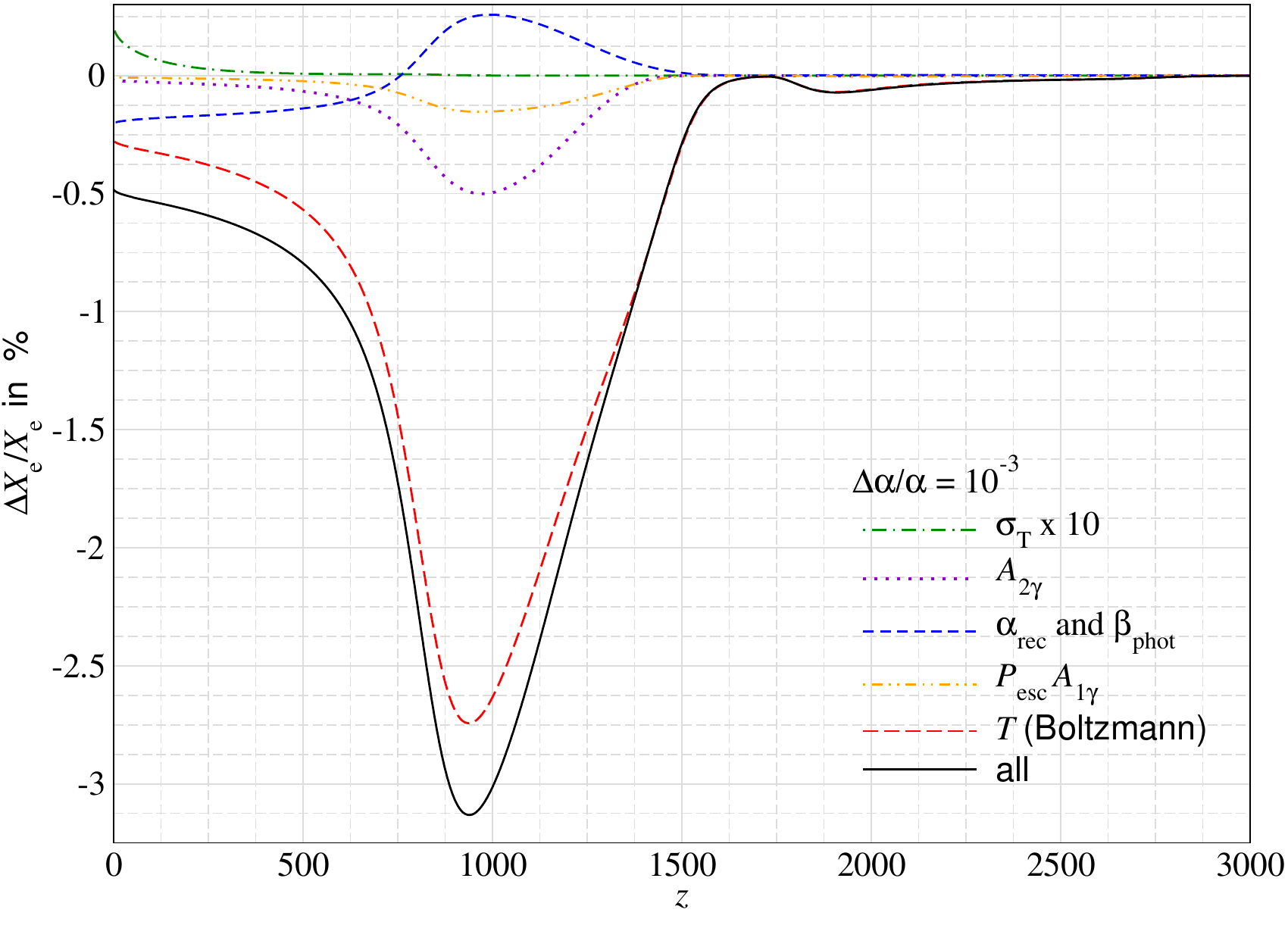}
    \caption{The relative changes in the ionization history for $\dalpha=10^{-3}$ with respect to the standard case caused by different effects. {\tt Recfast++} was used for the computations. The rescaling of temperature ($\leftrightarrow$ mainly affecting the Boltzmann factors) yields $\Delta X_{\rm e}/X_{\rm e}\simeq -2.7\%$, dominating the total contributions, which peaks with $\simeq -3.1\%$ at $z\simeq 1000$. Note that the modification due to $\sigma_{\rm T}$ has been scaled by a factor of ten to make it visible. The figure was taken from \cite{Hart2017}.}
    \label{fig:cont}
\end{figure}
The relative changes to the ionization history, $\Delta X_{\rm e}/X_{\rm e}$, for the different quantities in Eq.~\eqref{eq:recfast_scaling} are illustrated in Fig.~\ref{fig:cont}. We chose a value for $\dalpha=10^{-3}$, which leads to a percent-level effect on $X_{\rm e}$.
As expected, the biggest effect appears after rescaling the temperature for the evaluation of the photonionization and recombination rates. More explicitly, this can be understood when considering the net recombination rate to the second shell, which can be written as\footnote{In full equilibrium, $\Delta R_{\rm con}=0$.} $$\Delta R_{\rm con}= \Ne N_{\rm p} \alpha_{\rm rec}-N_{\rm 2}\,\beta_{\rm phot}=\alpha_{\rm rec}[\Ne N_{\rm p}-g(\Tg)N_{\rm 2}],$$ where $g(\Tg)\propto \Tg^{3/2}\,\expf{-h\nu_{\rm 2c}/k\Tg}$ with continuum threshold energy, $E_{\rm 2c}=h\nu_{\rm 2c}$.
Here, the exponential factor ($\leftrightarrow$ Boltzmann factor) is most important, leading to an exponential effect once we replace $\Tg'(z)=\Tg(z) \times(\aEM/\aEMs)^{-2}(\me/\mes)^{-1}$. 

The second largest individual effect is due to the rescaling of the two-photon decay rate, $A_{2\gamma}$. This is expected since $\aEM$ appears in a high power, $A_{2\gamma}\propto \aEM^8$, and also because the 2s-1s two-photon channel plays such a crucial role for the recombination dynamics \cite{Zeldovich68, Peebles68, Chluba2006}, allowing $\simeq 58\%$ of all hydrogen atoms to become neutral through this route \cite{Chluba2006b}. 

The normalizations of the recombination and photoionization rates (blue/dashed line in Fig.~\ref{fig:cont}) give rise to a net delay of $\Delta X_{\rm e}/X_{\rm e}\simeq 0.3\%$ at $z\simeq1000$, which partially cancels the acceleration due to $A_{2\gamma}$. This is due to the stronger scaling of $\beta_{\rm phot}$ with $\aEM$ than $\alpha_{\rm rec}$. At low redshifts ($z\lesssim 750$), recombination is again accelerated, indicating that a higher fraction of recombination events occurs, as the importance of photoionization ceases. 
The correction related to the Lyman-$\alpha$ channel is found to be $\simeq 3.3$ times smaller than for the two-photon channel, yielding $\Delta X_{\rm e}/X_{\rm e}\simeq -0.15\%$ at $z\simeq1000$.
Finally, the effect of rescaling $\sigma_{\rm T}$ are very small and only becomes noticeable at low redshifts. At these redshifts, the matter and radiation temperature begins to depart, giving $\Te < \Tg$. For larger $\aEM$, this departure is delayed, such that $\Te$ stays longer close to $\Tg$. Hotter electrons recombine less efficiently, so that a slight delay of recombination appears (cf., Fig.~\ref{fig:cont}).
%

\begin{figure}
  \centering
 \includegraphics[width=0.85\columnwidth]{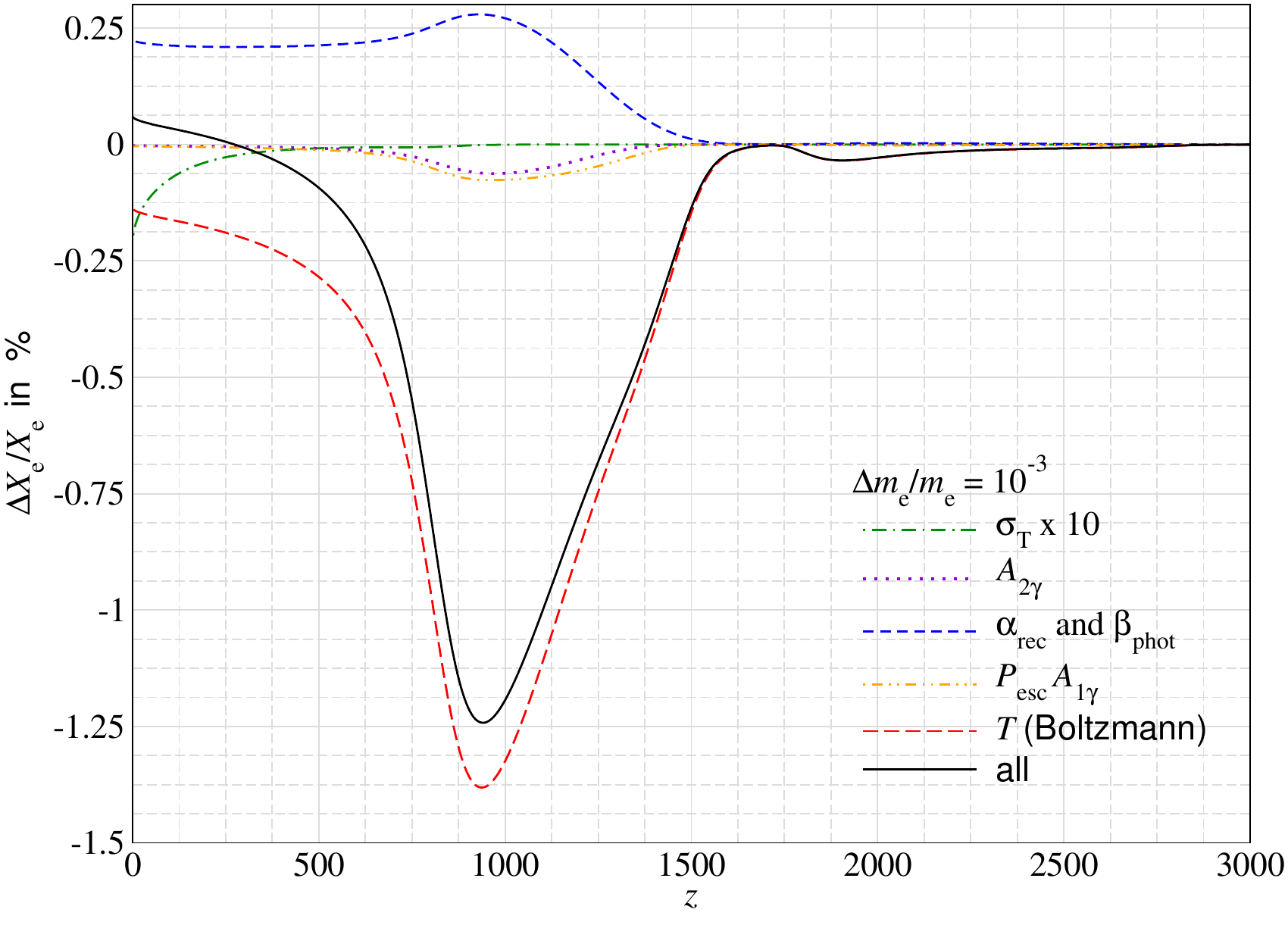}
  \caption{Same as in Fig.~\ref{fig:cont} but for $\dme=10^{-3}$. The effective temperature rescaling again dominates the total change. Around $z\simeq 1000$, the total effect is $\simeq 2.5$ times smaller than for $\dalpha=10^{-3}$ mainly due to the weaker dependence of the level energy on $\me$. The figure was taken from \cite{Hart2017}.}
  \label{fig:cont_me}
\end{figure}

\vspace{-0mm}
\subsection{Ionization history modifications due to variation of $\me$}
\label{sec:Xe_me}
We now briefly illustrate the changes caused by the effective electron mass, with the variation parametrized as $\me=\mes(1+\dme)$ with respect to the standard value $\mes$. Inspecting the scalings of Eq.~\eqref{eq:recfast_scaling}, one expects the overall effect to be smaller than for $\aEM$, but otherwise very comparable. For example, the effect of temperature rescaling should be roughly half as large. Similarly, the effect due to rescaling $A_{2\gamma}$ should be roughly $8$ times smaller, and so on. This is in good agreement with our findings (cf. Fig.~\ref{fig:cont_me}). 

The net effect on $X_{\rm e}$ is about 2.5 times smaller than for $\aEM$ around $z\simeq 1000$. This suggests that the CMB constraint on $\me$ is weakened by a similar factor. However, adding the rescaling of the Thomson cross section for the computation of the visibility function strongly enhances the geometric degeneracy for $\me$, such that the CMB only constraint on $\me$ is $\gtrsim 20$ times weaker than for $\aEM$ (see Sect.~\ref{sec:CMB_constraints}). In addition, a small difference related to the renormalizations of the photoionization and recombination rates (blue/dashed line) appears. For $\dme>0$, the photoionization rate is increased and the recombination rate is reduced for these contributions [cf. Eq.~\eqref{eq:recfast_scaling}]. Both effects delay recombination (see Fig.~\ref{fig:cont_me}). Thus, around $z\simeq 10^3$ the net effect is slightly larger than for $\aEM$. In contrast to $\aEM$, at late times no net acceleration of recombination occurs. These effects slightly modify the overall redshift dependence of the total $X_{\rm e}$ change, in addition to lowering the effect in the freeze-out tail. 
At the level $\dme\simeq 1\%$, additional higher order terms become important, allowing one to break the near degeneracy between $\aEM$ and $\me$ in joint analyses \cite{Planck2015var_alp}. At the level of the current CMB only constraints, this aspect indeed is of relevance (Sect.~\ref{sec:CL_results}).

\subsection{Comparison with {\tt CosmoRec} and generalized VFC models.}
\label{sec:CosmoRec_beyond}
We highlight that \cite{Hart2017} also explicitly demonstrated that a full treatment of the problem using the advanced recombination code {\tt CosmoRec} yields results that are very similar to those from a simpler three-level treatment with correction function approach \cite{Shaw2010} to mimic the recombination physics corrections. This also allowed \cite{Hart2017} to perform calculations with explicit time-dependence of $\aEM$ and $\me$, initially focusing on a simple power-law redshift dependence
\begin{equation}
\label{eq:powerlaw}
  \aEM(z) = \aEM(z_0)\,\left(\frac{1+z}{1100}\right)^p,
\end{equation}
for $\aEM$ and similarly for $\me$.
Using a principal component analysis (PCA) \cite{Hart2022VFCPCA}, which extended the previous recombination perturbation framework developed in \cite{Farhang2011, Farhang2013, Hart2020PCA}, general VFC variations around the recombination era were further studied. This showed that time-dependence can in fact be independently constrained already with existing data and also led to various generalized limits on VFCs during recombination \cite{Lee2023VFC, Tohfa2023}, which we will briefly highlight below.

\section{CMB anisotropy constraints}
\label{sec:CMB_constraints}

\subsection{Propagating the effects to the CMB anisotropies}
\label{sec:CL_calc}
The temperature and polarization power spectra of the CMB depend on the dynamics of recombination through the ionization history, which defines the Thomson visibility function and last scattering surface \cite{Sunyaev1970, Peebles1970, Hu1996anasmall}. Therefore, variations of $\aEM$ and $\me$ can leave a direct imprint on the CMB power spectra. A detailed description of changes to the visibility function and various additional illustrations also for time-dependent VFCs can be found in \cite{Hart2017, Hart2022VFCPCA}. Here, we highlight the changes to the CMB temperature power spectra for variations of $\aEM$ and $\me$, noting that those in polarization show very similar features. The power spectra were computed using {\tt CAMB} \cite{CAMB} for the standard cosmology \cite{Planck2015params}. 

To propagate the effect of VFCs to the CMB anisotropies, two changes are required. The standard recombination history has to be replaced as explained in the previous section. In addition, the Thomson scattering rate has be to updated in the Boltzmann code using the modified Thomson cross section. For the latter, two approaches are possible, one based on directly modifying the Boltzmann code, the other on mimicking the effect by rescaling the ionization history. Both seem to deliver consistent final results \cite{Hart2017}. However, we stress how important it is to include the changes to the Thomson scattering rate, as without this effect $\aEM$ and $\me$ variations essentially exhibit a very similar phenomenology \cite{Hart2017, Hart2020Hubble}.

\begin{figure}[t]
	\centering
 \includegraphics[width=0.85\columnwidth]{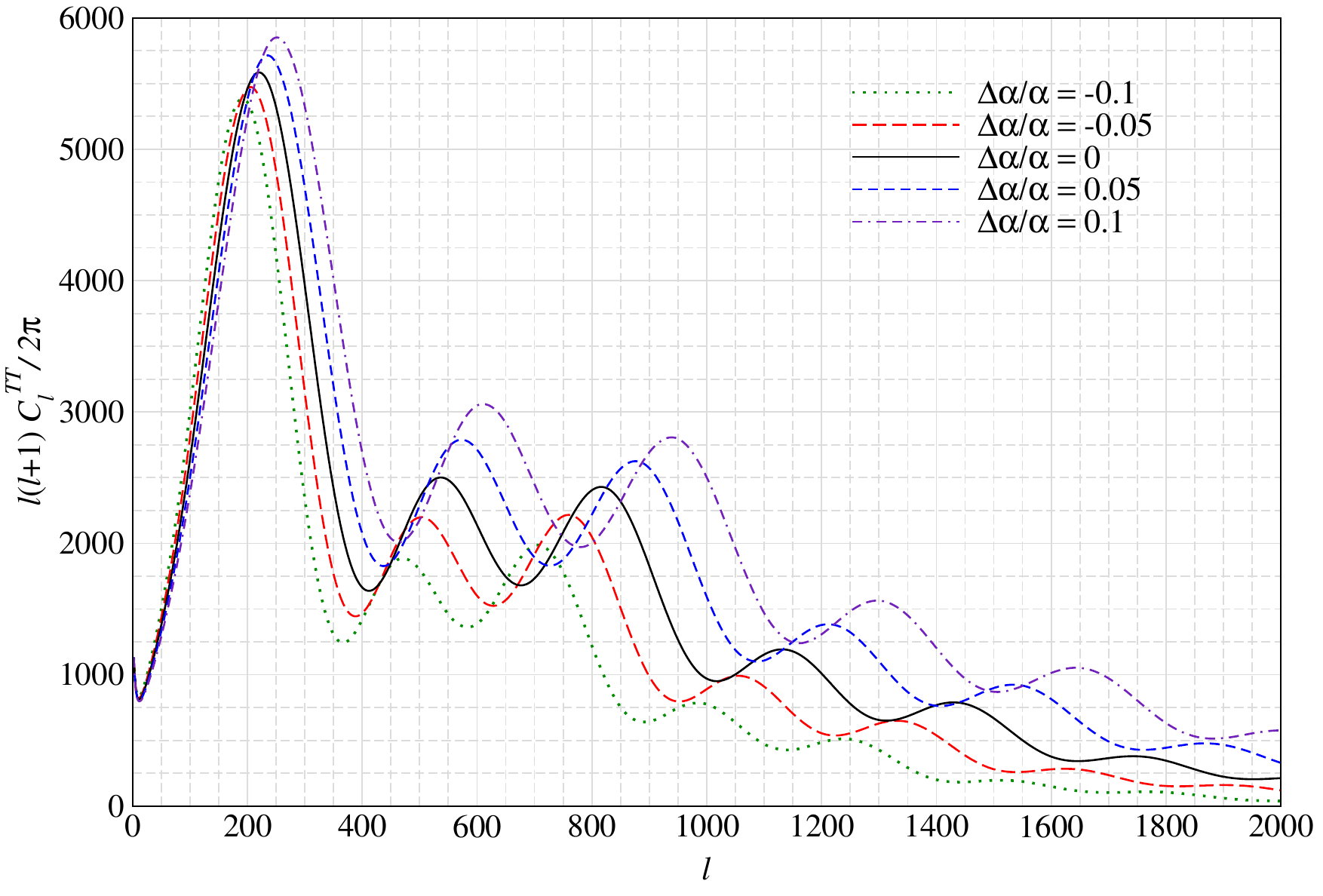}
    \caption{The CMB temperature power spectra for different values of $\aEM$. This shows that as the fine structure constant increases, the anisotropies shift toward smaller scales and higher amplitudes. The figure was taken from \cite{Hart2017}.}
    \label{fig:alpha_cl}
\end{figure}
In Fig.~\ref{fig:alpha_cl}, we illustrate the effect of $\aEM$ on the CMB temperature power spectrum. Two main features are visible: Firstly, the peaks of the power spectrum are shifted to smaller scales (larger $\ell$) when $\dalpha >0$. This happens because earlier recombination moves the last scattering surface towards higher redshifts, which decreases the sound horizon and increases the angular diameter distance to recombination \cite{Kaplinghat1999, Battye2001}. 
Secondly, for $\dalpha>0$, the peak amplitudes are enhanced. This is mainly because earlier recombination suppresses the effect of photon diffusion damping on the anisotropies \cite{Kaplinghat1999, Battye2001}.
For variations of $\me$, very similar responses are found, but with an amplitude that is reduced by a factor of $\simeq 2-3$ \cite{Hart2017}.

\begin{figure}
  \centering
  \includegraphics[width=0.85\linewidth]{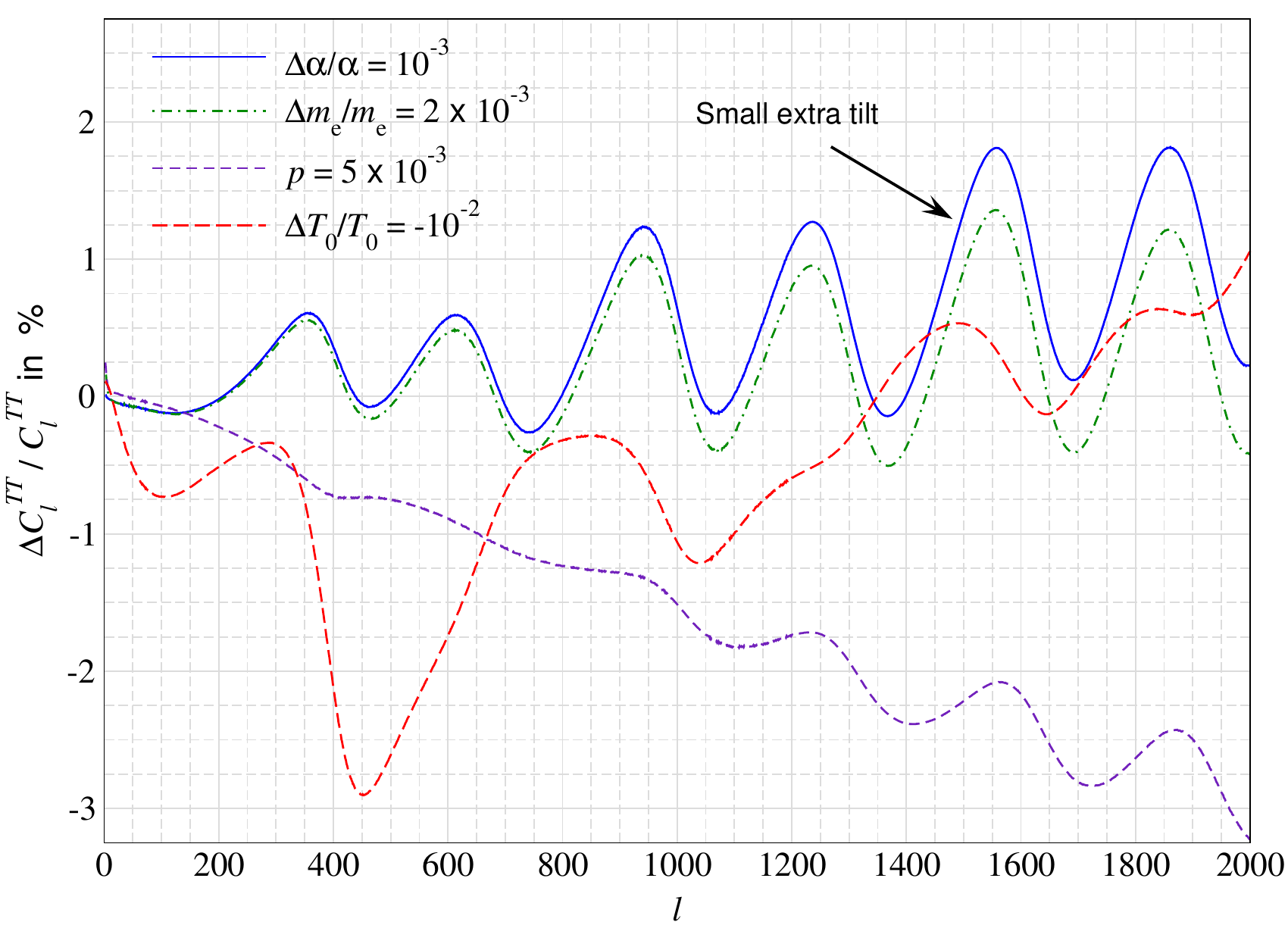}
  \caption{Comparison of the CMB $TT$ power spectrum deviations when varying $\aEM$, $\me$, $T_0$ and $p$. We chose $\dalpha=10^{-3}$, $\dme=2\times 10^{-3}$, $\Delta T/T=-10^{-2}$ and $p=5\times 10^{-3}$ (simultaneously for $\aEM$ and $\me$) to obtain effects at a similar level. Notice the small extra tilt when comparing the case for $\aEM$ with $\me$, which helps when constraining $\aEM$. The figure was taken from \cite{Hart2017}.}
  \label{fig:talpha}
\end{figure}
For small $\dalpha$ and $\dme$, we illustrate the relative change of the temperature power spectrum in Fig.~\ref{fig:talpha}. The effect on the peak positions is more noticeable than the small overall tilt caused by changes related to diffusion damping. As expected, the changes to the CMB $TT$ power spectra, $\Delta C_{\ell}/C_{\ell}(\dalpha)$ and $\Delta C_{\ell}/C_{\ell} (\dme)$, become almost indistinguishable when using $\dme\approx (2-3)\,\dalpha$.
This presents a quasi-degeneracy between the two parameters and also suggests that naively the analysis for $\dalpha$ could be sufficient to estimate the errors for a corresponding analysis of $\me$. 
However, when constraining $\me$, an enhanced geometric degeneracy, because of the differing effect of $\sigT$, inflates the error to the percent level \cite{Planck2015var_alp, Hart2017}. In this case, higher order terms become important and the near degeneracy is broken. When also adding information from BAO, the error on $\me$ is strongly reduced, and a simple scaling of the errors, $\sigma(\dme)\simeq 3\sigma(\dalpha)$, is recovered \cite{Hart2017}.

In Fig.~\ref{fig:talpha}, we also illustrate the effect of varying the average CMB temperature, $T_0$, and a time-dependent model for both $\aEM$ and $\me$ with a phenomenological power-law as in Eq.~\eqref{eq:powerlaw} around a pivot redshift of $z=1100$. Changes in the CMB monopole temperature show a different response pattern than VFCs, making these two effects principally distinguishable. This is because varying the CMB temperature affects the ionization history (at leading order like VFCs), but without changing the Thomson scattering rate. In addition, $T_0$
modifies the matter-radiation equality and therefore has a separate overall effect. Similarly, we conclude that the power-law index of the time-dependent model can be independently constrained \cite{Hart2017}. 

\subsection{Constraints from \Planck}
\label{sec:CL_results}
Now that we have developed a detailed understanding about how the CMB anisotropies are affected by changes of $\aEM$ and $\me$, we can directly consider some of the existing constraints from \Planck. Early constraints were derived using \Planck 2013 data in \cite{Planck2015var_alp}. Aside from additional data (e.g., \Planck polarization) and improvements in the understanding of systematics and calibration, the later analysis of the \Planck 2015 data yielded similar constraints \cite{Hart2017}, although with slightly improved errors.\footnote{For a detailed discussion of the effects of various analysis choices we refer to \cite{Hart2017}.} 
Here we highlight the latest \Planck 2018 results, which were obtained in \cite{Hart2020Hubble}. For details we refer the interested reader to that paper.

\begin{table*}[t]
\centering
\begin{tabular} { l c | c c}
\hline
\hline
Parameter &\Planck 2018 &\Planck 2018 &\Planck 2018 
\\
 & &  + varying $\aEM$ & + varying $\me$ 
\\
\hline
$\Omega_b h^2  $ &  $0.02237\pm 0.00015  $ &  $0.02236\pm 0.00015  $ &  $0.0199^{+0.0012}_{-0.0014}  $ 
\\
$\Omega_c h^2  $ &  $0.1199\pm 0.0012  $ &  $0.1201\pm 0.0014  $ &  $0.1058\pm 0.0076$ 
\\
$100\theta_{MC}  $ &  $1.04088\pm 0.00031  $ &  $1.0416\pm 0.0034  $ &  $0.958\pm 0.045  $ 
\\
$\tau  $ &  $0.0542\pm 0.0074  $ &  $0.0540\pm 0.0075  $ &  $0.0512\pm 0.0077  $ 
\\
${\rm{ln}}(10^{10} A_{\rm s})  $ &  $3.044\pm 0.014  $ &  $3.043\pm 0.015  $ &  $3.029\pm 0.017 $ 
\\
$n_{\rm s}  $ &  $0.9649\pm 0.0041  $ &  $0.9637\pm 0.0070  $ &  $0.9640\pm 0.0040  $ 
\\
\hline
$\alpha_{\rm EM}/\alpha_{\rm EM,0}  $ & $--$   &  $1.0005\pm 0.0024  $ & $--$   
\\
$m_{\rm e}/m_{\rm e\,,0}  $ & $--$   & $--$ &  $0.888\pm 0.059  $ 
\\
\hline
$H_0 \,[{\rm km \, s^{-1}\,\Mpc^{-1}}] $ &  $67.36\pm 0.54  $ &  $67.56\pm 0.99  $ &  $46^{+9}_{-10}  $ 
\\
\hline\hline
\end{tabular}
\caption{Marginalised values of the fine structure constant and effective electron mass $\aEM$ and $\me$ using the \Planck 2018 data alone. A very wide prior $\left(H_0>20\,{\rm km \, s^{-1}\,\Mpc^{-1}}\right)$ for $H_0$ was used to avoid biasing the marginalised $\me$ posterior, which affected some of the results of the 2013 analysis \cite{Hart2017, Hart2020Hubble}.}
    \label{tab:Planck_alpha_me2018}
\end{table*}

\begin{figure}[t]
    \centering
    \includegraphics[width=0.95\linewidth]{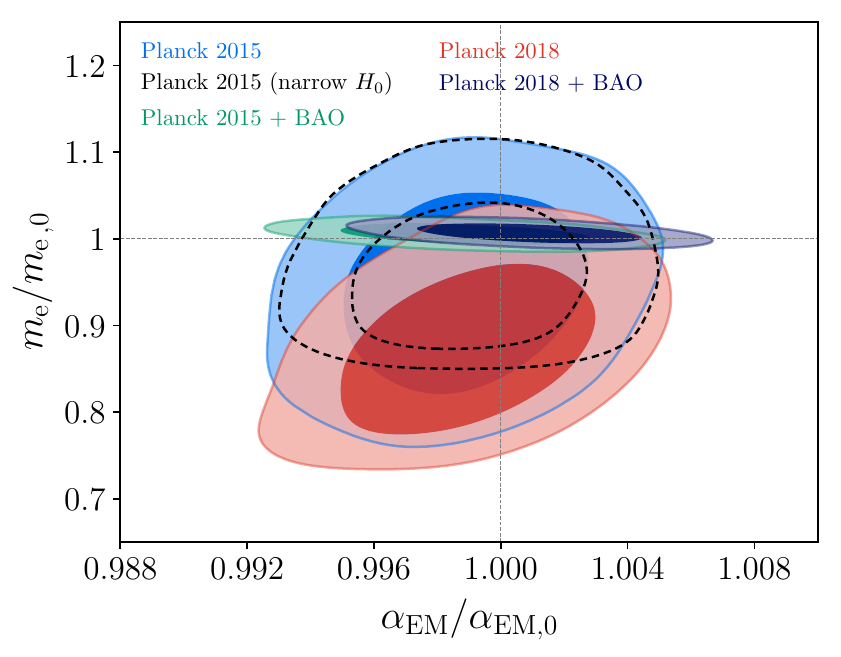}
    \caption{Posterior contours between $\aEM$ and $\me$ for the \Planck 2015 and 2018 data along with BAO contributions. Note that the dashed contour shows the 2015 contour but with a tighter prior on $H_0\in \{40,100\}\,{\rm km \, s^{-1}\,\Mpc^{-1}}$ to conform with the previous \Planck 2013 analysis. The figure was taken from \cite{Hart2020Hubble}.}
    \label{fig:malpha_mva}
\end{figure}

In Table~\ref{tab:Planck_alpha_me2018}, we summarize some of the \Planck 2018 constraints on $\aEM$ and $\me$, including the 2018 baseline \Planck data, with low-$\ell$ and high-$\ell$ data for temperature and $E$-mode polarisation power spectra, along with the lensing data from the same release \cite{Planck2018like,Planck2018Lensing}. The addition of $\aEM$ marginally affects the values and errors of the six standard parameters, yielding $\alpha_{\rm EM}/\alpha_{\rm EM,0}=1.0005\pm 0.0024$. In contrast, varying $\me$ has a strong effect especially on $H_0$, shifting it to extremely low values and allowing for a low value of the electron rest mass, $\me/\mes = 0.888\pm 0.059$. This is caused by the small differences in the way $\me$ affects the CMB power spectra, with a crucial role played by adding changes to $\sigT$ \cite{Hart2017, Hart2020Hubble}. As we discuss below, this large geometric degeneracy is one of the key ingredients for alleviating the Hubble tension when combined with supernova data.

In Fig.~\ref{fig:malpha_mva}, we show the two-dimensional posteriors for the constraint on simultaneously varying values of $\aEM$ and $\me$. For comparison, the contours relating to the 2015 analysis and also combinations with baryon acoustic oscillation (BAO) \cite{SDSSDR12} data are illustrated. 
The constraints for \Planck alone exhibit very wide posteriors, with slight differences in the centroids of the 2015 and 2018 constraints. We note that a narrower prior of $H_0>40\, {\rm km \, s^{-1}\,\Mpc^{-1}}$ (conforming with the initial CMB analysis of \Planck\cite{Planck2015var_alp}) affects the posterior for the 2015 data, shrinking it in the $\me$-direction. This further highlights the significant differences in the role of $\aEM$ and $\me$ on the CMB anisotropies.

The introduction of BAO data significantly tightens the constraints on $\me$ and we can also observe a small drift in the centeral value of $\aEM$. The obtained value changes from $\aEM/\aEMs=0.9989\pm0.0026$ for the 2015 data to $\aEM/\aEMs=1.0010\pm0.0024$ with 2018 data, both with BAO included and when simultaneously varying $\me$. By contrast, there is no drift for $\me$, with $\me/\mes=1.0056\pm0.0080$ changing to $\me/\mes=1.0054\pm0.0080$ for \Planck 2015+BAO and \Planck 2018+BAO, respectively. 
These findings highlight a strong level of agreement of \Planck with BAO data, which has been emphasized on many occasions \cite{Planck2018params}.

\begin{figure*}
    \centering
    \includegraphics[width=.87\linewidth]{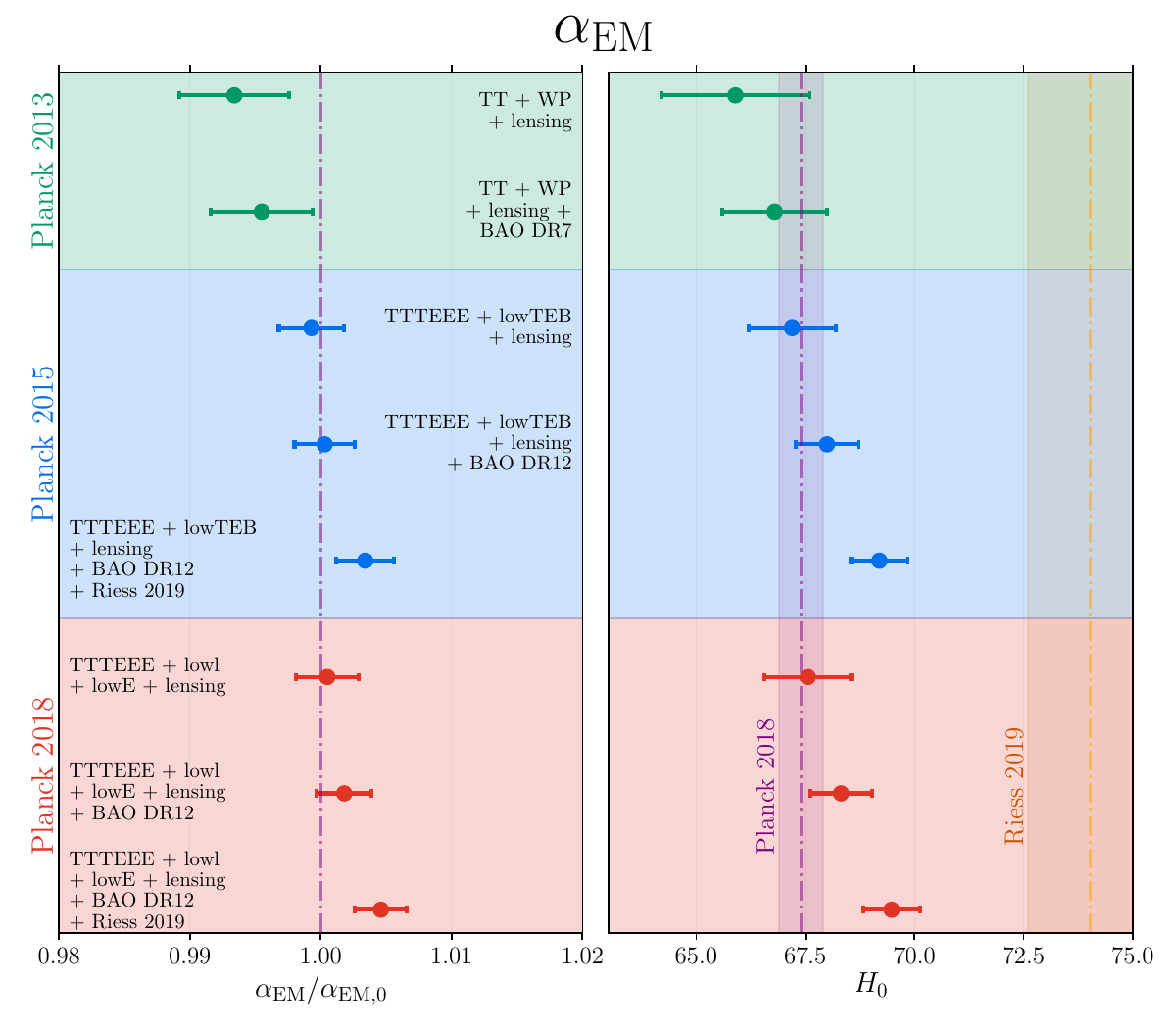}
    \centering
    \includegraphics[width=.87\linewidth]{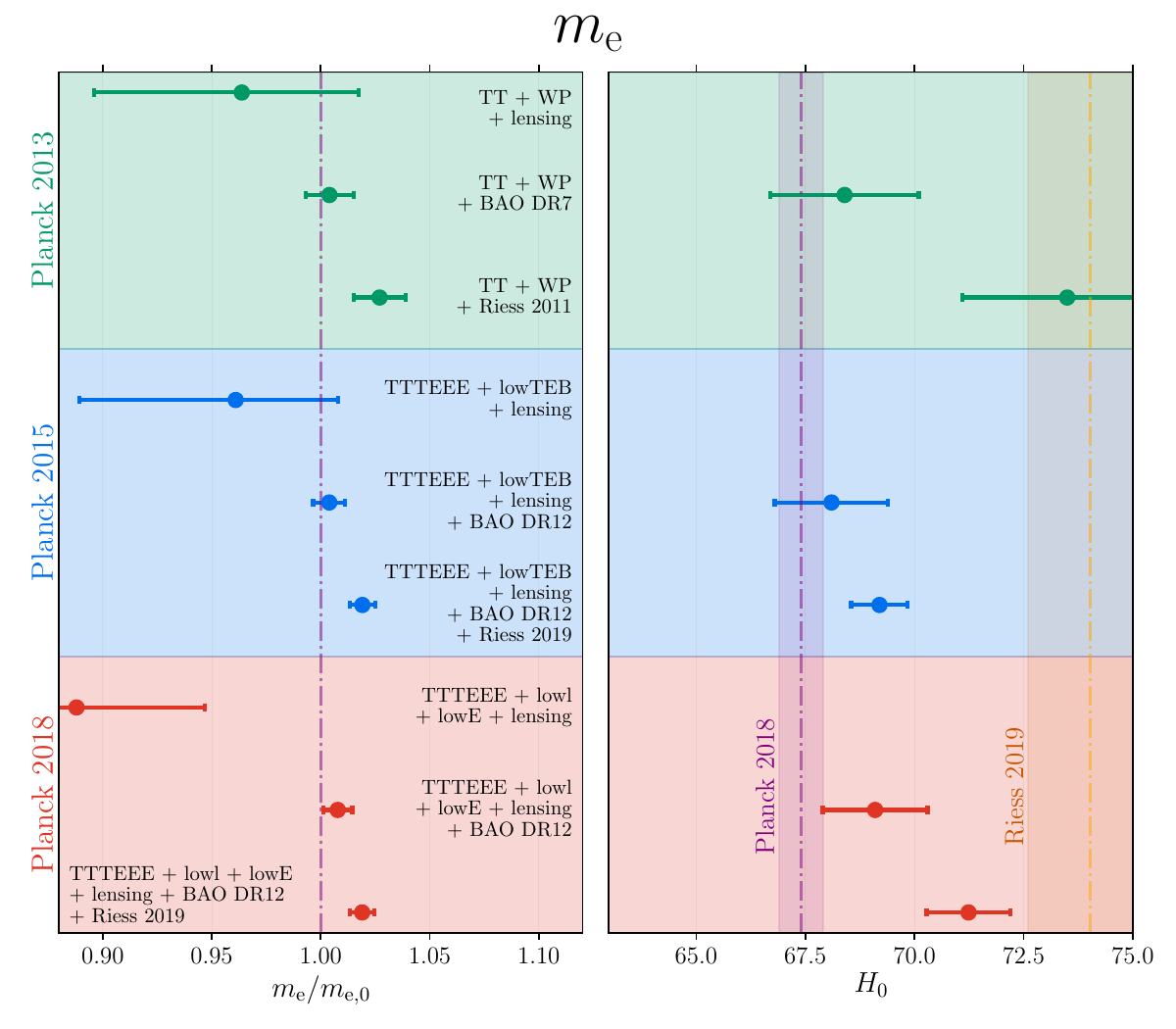}
    \caption{Constraints on the fundamental constants ({\it left}) using various combinations of \Planck data together with their $H_0$ values and errors ({\it right}). 
    %
    %
     {\it Top:} results from the fine structure constant $\aEM$. {\it Bottom:} similar results but from the effective electron mass $\me$. Here, we have redacted the constraint for $H_0$ from CMB data only because the error bars are so large. For the $\me$ MCMC analysis, we have widened the prior on the Hubble constant such that $H_0>20\,{\rm km \, s^{-1}\,\Mpc^{-1}}$. Figure is from \cite{Hart2020Hubble}.}
    \label{fig:me_alpha}
\end{figure*}

\section{Alleviating the Hubble tension with VFCs}
\label{sec:Hubble}
The attentive reader will already have noticed the route forward to alleviating the Hubble tension through varying the electron rest mass. This possibility was first noticed in \cite{Hart2020Hubble}, where the constraints on VFCs from different data combinations were studied, also adding supernova/Cepheid (R19) \cite{Riess2019} data. 

A summary of the constraints is shown in Fig.~\ref{fig:me_alpha} with particular focus on the interplay with $H_0$.
For $\aEM$, we can notice broadly consistent constraints across all data combinations with only a small shift in the value of $H_0$ towards R19 in the combined constraints, indicating a resistance in terms of geometric freedom. For $\me$, two effects are found. First, as pointed out above, when adding BAO data, the error on $\me$ is significantly reduced in comparison to the CMB only constraints, bringing $H_0$ into agreement with the only CMB inference. Second, when also adding R19, a non-standard value of $\me$ (at $\simeq 3.5\sigma$ significance) is traded for a reduction of the Hubble tension. This begs the question whether $\me$ could indeed play a role in this problem. A model comparison study carried out in \cite{NilsH0Olympics} indeed indicated that a simple variation of $\me$ provides a good contender in this respect, although not all issues could be resolved.

\subsection{Adding curvature}
\label{sec:curvature}
As Fig.~\ref{fig:me_alpha} clearly shows, a constant variation of $\me$ does not fully solve the Hubble tension, albeit reducing it below $2\sigma$. As one possible extension, \cite{Hart2020Hubble} also studied solutions with power-law VFC time-dependence, but found this to not further improve matters. A little later, \cite{Sekiguchi2021PhRvD} studied cosmologies with non-zero curvature in addition to variations of $\me$, finding these models to solve the Hubble tension. Indeed, this possibility was the winning finalist in the $H_0$-Olympics model comparison exercise \cite{NilsH0Olympics}. However, allowing for non-zero curvature does open yet another non-standard direction in cosmology (in addition to accepting varying $\me$), with strong resistance in terms of standard inflationary predictions \cite{DiDio2016}. One could feel inclined to cling on to zero curvature cosmologies given the great successes of the inflationary paradigm, but ultimately observations and careful analysis will have to decide.

\begin{figure}
    \centering
    \includegraphics[width=0.95\linewidth]{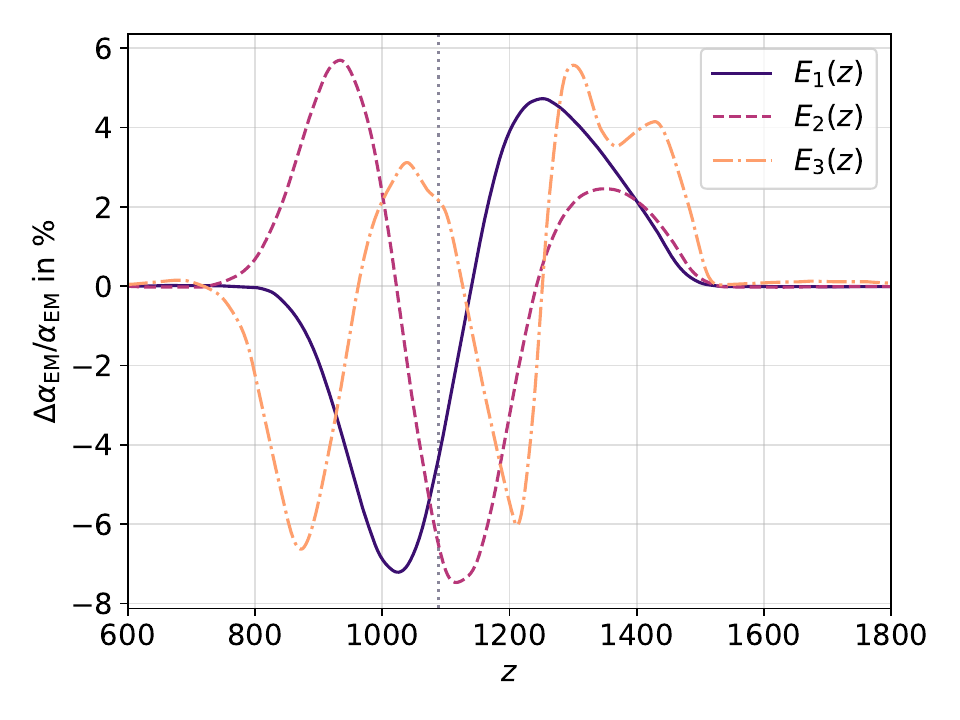}
    \includegraphics[width=0.95\linewidth]{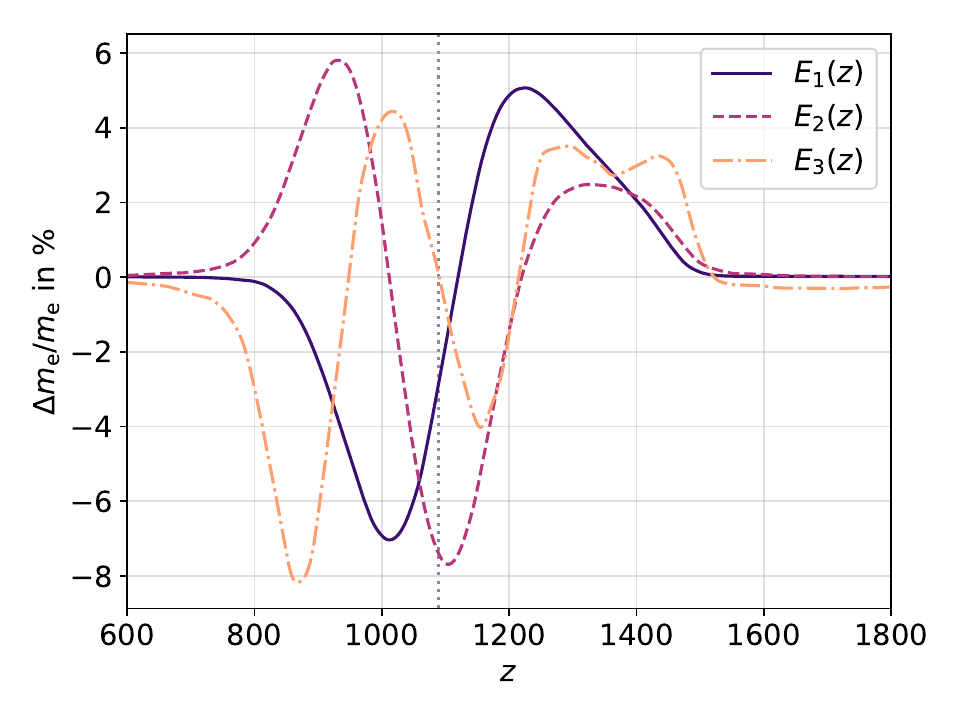}
    \caption{The first three VFC principal components for $\aEM$ \emph{(top)} and $\me$ \emph{(bottom)} as constructed using \Planck 2018 data. The eigenmodes are all normalised such that $\int|E_i^2(z)|\id z = 1$. The maximum of the $\Lambda{\rm CDM}$ Thomson visibility function has been marked as vertical dotted line. Figure is taken from \cite{Hart2022VFCPCA}.}
    \label{fig:modesPlanck}
\end{figure}

\subsection{Time-dependent VFCs models}
\label{sec:PCAVFC}
Most of the results presented above assumed constant (i.e., time-independent/single-valued) changes to the values of FCs in the early Universe. As highlighted in \cite{Hart2017}, explicitly time-dependent VFCs can in principle be constrained independently even with existing CMB data, given that the CMB responses are distinct (see Fig.~\ref{fig:talpha}). This idea was later generalized by applying a principal component analysis (PCA) to possible time-dependent VFC perturbations around the standard value in the hopes to further diminish the Hubble tension \cite{Hart2022VFCPCA}. The idea is very simple: given the (CMB) observables, it is difficult to limit a specific change in the values of FCs at a given redshift. However, certain correlated variations/perturbations (i.e., VFC modes) across wider ranges of redshifts can indeed be analysized and constrained. To construct these VFC eigenmodes, one can use the Fisher information matrix to assess the observability. This yields a ranked system of possible VFC modes that can be best constrained by the data. 
Similar approaches were applied to studies of perturbations in the reionization \cite{Mortonson:2009qv} and recombination \cite{Farhang2011, Farhang2013, Hart2020PCA} histories.

The first few $\aEM$ and $\me$ eigenmodes are shown in Fig.~\ref{fig:modesPlanck}. In a perturbative sense, most of the information on VFCs can be gleaned from around the maximum of the Thomson visiblity at $z\simeq 1100$. This is reflected in the localization of the VFC modes around this redshift. Applying these VFC modes to the \Planck 2018 data no significant detection of non-zero mode amplitudes was found \cite{Hart2022VFCPCA}. Similarly, when combining with BAO and R19, no additional improvement with respect to the Hubble tension over a constant variation was identified \cite{Hart2022VFCPCA}. However, to exclude the possibility of a more general time-dependent VFC history, this would require further investigations of time-dependent models that interface between the standard candle, low-redshift era and the surface of last scattering.

The PCA approach inherently assumes a perturbative variation of the FCs. This assumption need not hold, and in addition higher order eigenmodes that individually fall below the detection threshold (due to their more rapid redshift-variablity) could together allow for more general modifications \cite{Farhang2013}. In \cite{Lee2023VFC}, an assessment of which time-dependent change to $\aEM$ and $\me$ would be required to reduce the Hubble tension was carried out. It was demonstrated that more general time-dependent variations of $\aEM$ and $\me$ around recombination can solve the Hubble tension (and even reduce the $S_8$ tension) when applied to CMB data alone. However, once BAO and supernova data is added, full solutions to the Hubble tension evade a perturbative treatment, although extension of the framework to the non-perturbative regime can be done \cite{Lee2023VFC}. This points towards the possibility that more general theoretical models could indeed help to restore consistency between various probes, although more work is certainly needed.

\begin{figure}
    \centering
    \includegraphics[width=0.95\linewidth]{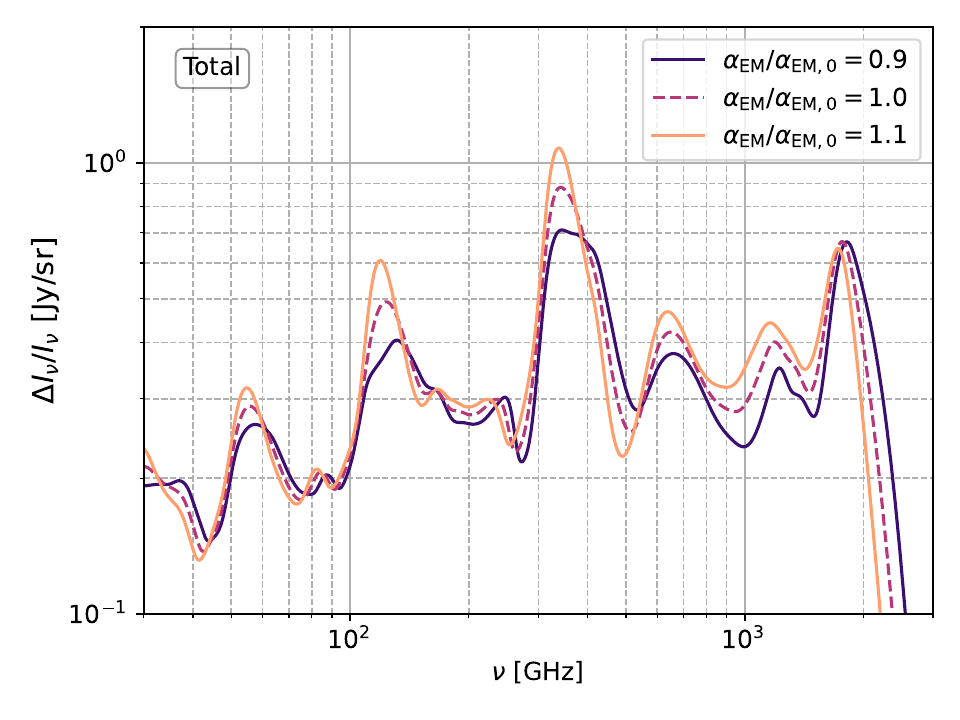}
    \includegraphics[width=0.95\linewidth]{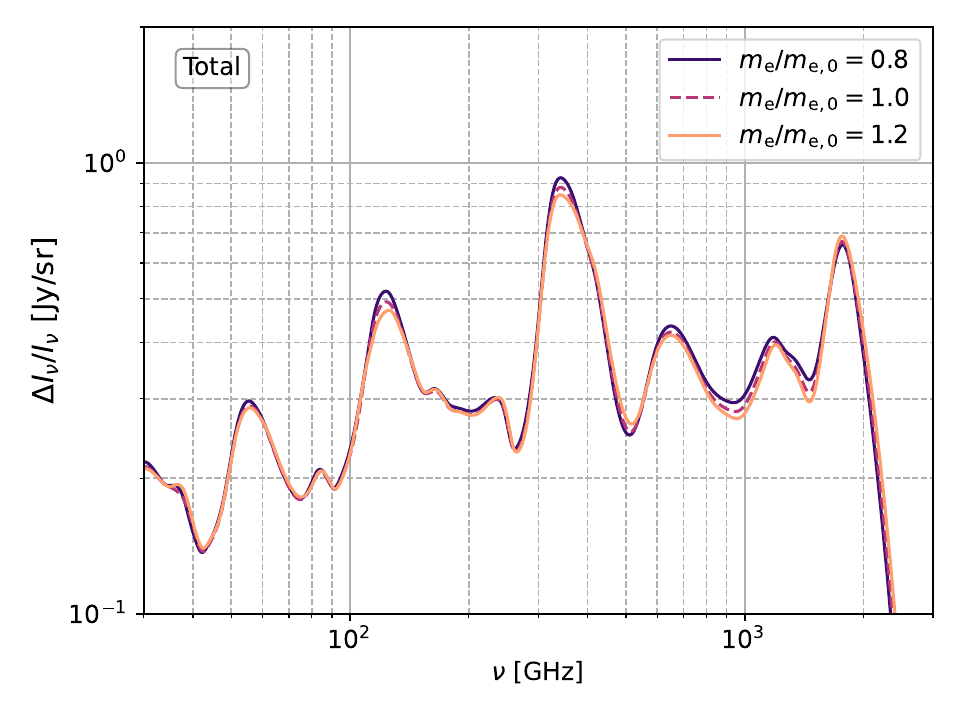}
    \caption{The total impact on the recombination lines from the variations for $\Delta\aEM/\aEMs=\pm0.1$ \emph{(top)}. The total variations due to changes in $\Delta\me/\mes=\pm0.2$ are shown for comparison \emph{(bottom)}. Figure taken from \cite{Hart2023VFCCRR}.}
    \label{fig:vfcCrrTotal}
\end{figure}
\begin{figure}
    \centering
    \includegraphics[width=\linewidth]{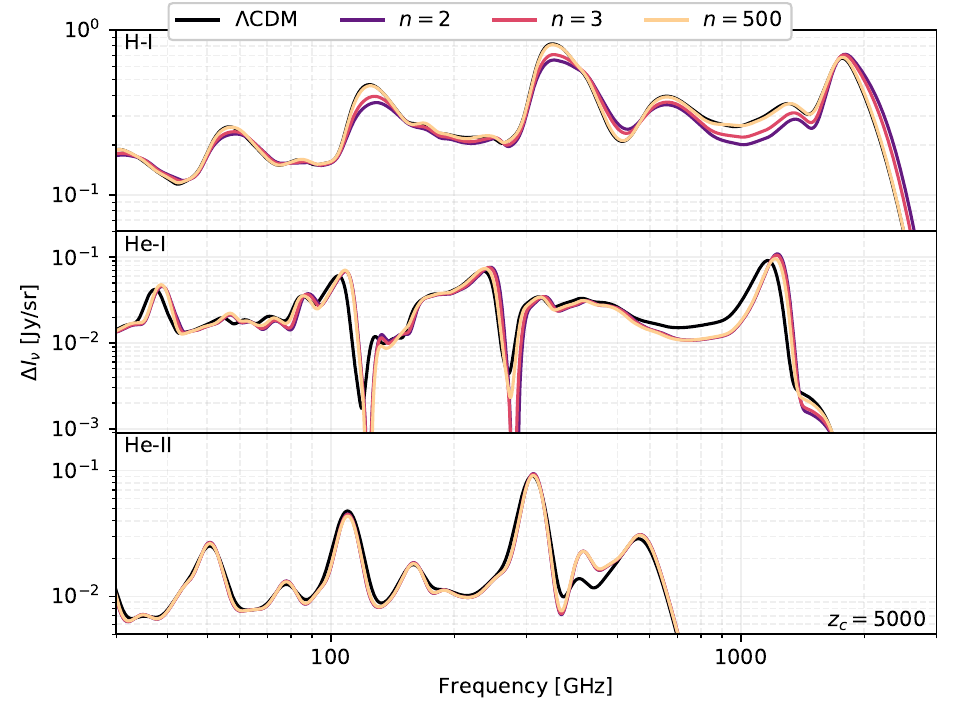}
    \caption{Comparison for an EDE model that has different slopes for $z_{\rm c}=5000$ across the different atomic species. Hydrogen (\emph{top}) is compared against the Helium I and II (\emph{middle,bottom}) lines, respectively. For better clarity, this has been plotted with $f_{\rm ede} = 0.8$. Figure taken from \cite{Hart2023VFCCRR}.}
    \label{fig:h_all}
\end{figure}
\begin{figure*}
    \centering
    \includegraphics[width=\columnwidth]{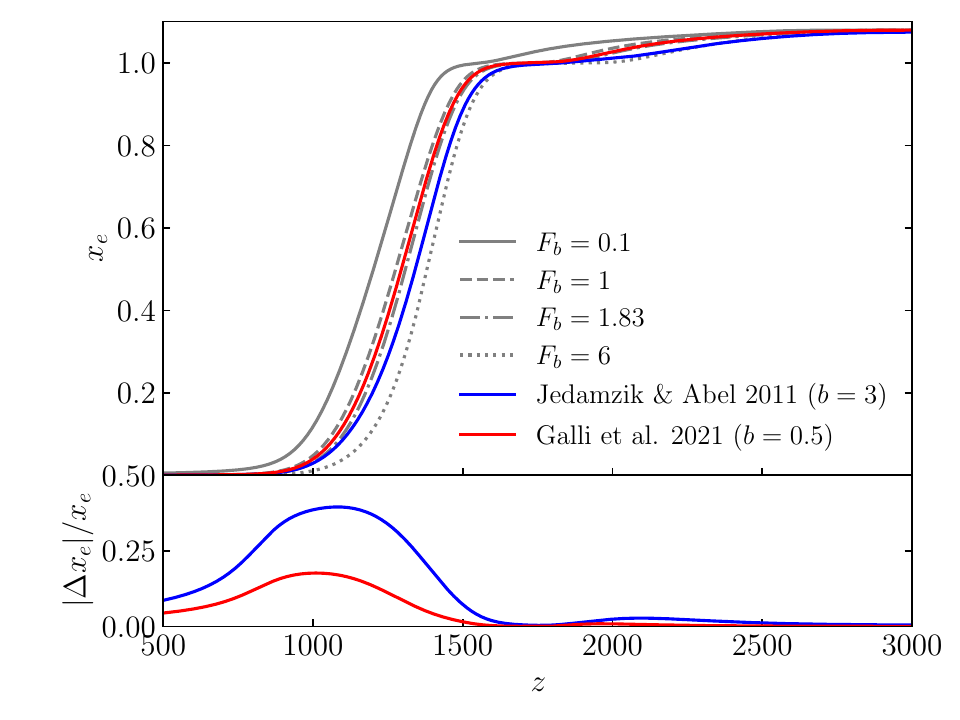}
    \includegraphics[width=\columnwidth]{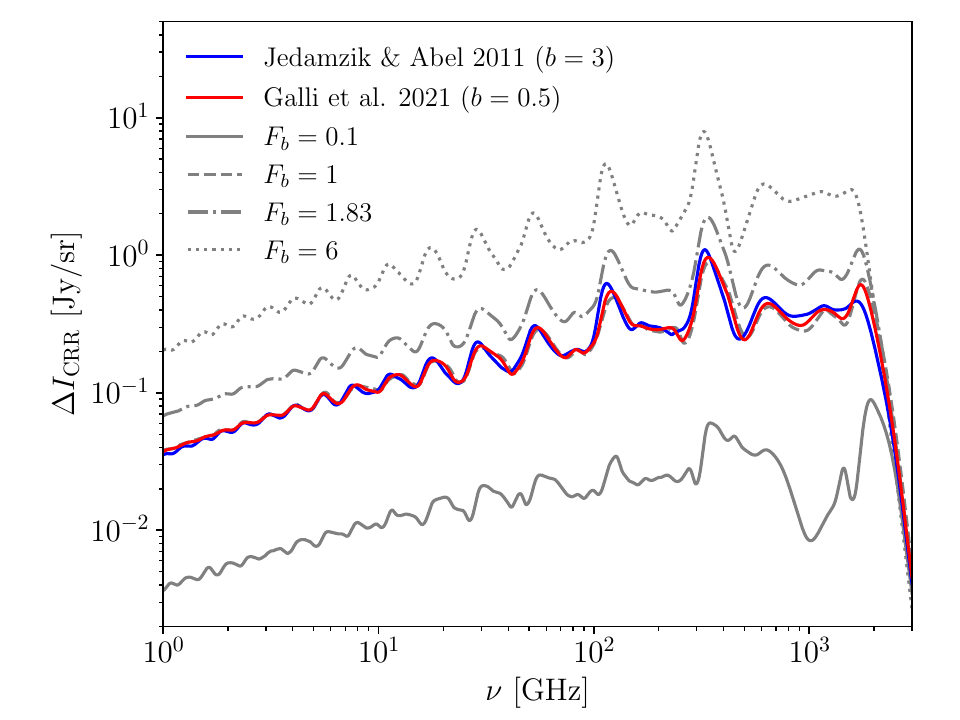}
    \caption{Recombination histories (upper panel) and CRR spectra (lower panel) for the PMF models similar to those consider in \cite{Jedamzik:2013gua, 
    Jedamzik2020Relieving, Galli:2021mxk}. The individual spectra for various values of the baryon density enhancement factor are shown as gray lines (the $F_{\rm b}=1$ case corresponds to \LCDM). The averages are instead displayed in blue and red, and show significant second order contributions, manifesting in smearing of the lines and shifts in their position.}
    \label{fig: PMFs}
\end{figure*}

\section{New insights from the cosmological recombination radiation}
\label{sec:CRR}
In the previous sections, we have primarily focused on illustrating the role of VFCs, and in particular of $\me$, in possible solutions to the Hubble tension. In this, one key ingredient is the associated modification to the cosmological recombination history, although it seems clear that a simple perturbative change does not provide a sufficient degree of freedom to fully resolve the tension. However, stepping back from a VFC driven solution, one important question is how we could tell if indeed a (strongly) modified recombination history is responsible for the Hubble tension? Furthermore, solutions based on PMFs (see \PMFC) are also mainly successful because of their modifications to the average recombination history. It is thus important to ask how could we distinguish different options?

One important avenue forward is to use more accurate measurements with upcoming CMB experiments such as The Simons Observatory \cite{SOWP2018} and CMB-S4 \cite{CMBS42016} to improve the constraints on VFC \cite{Hart2020Hubble, Hart2022VFCPCA}. However, ultimately access to new observables will be required.

The cosmological recombination processes is also associated with the emission of photons from the Hydrogen and Helium plasma \cite{Dubrovich1975, RybickiDell94, DubroVlad95, Burgin2003, Kholu2005, Jose2005, Chluba2006b, Jose2008}. The cosmological recombination radiation (CRR) can now be accurately computed using {\tt CosmoSpec} \cite{Chluba2016CosmoSpec}, which was extended to also incorporate the effect of VFCs \cite{Hart2023VFCCRR}. This CMB distortion signal may become observable \cite{Hart2020CRR} with future CMB spectrometers akin to {\it PIXIE} \cite{Kogut2011PIXIE, Kogut2016, Chluba2021Voyage}, opening the exciting possibility to directly study the dynamics of the recombination process \cite{Chluba2008T0, Sunyaev2009}.

In Fig.~\ref{fig:vfcCrrTotal}, we illustrate the effect of varying $\aEM$ and $\me$ on the CRR. Firstly, one can clearly see that for the chosen parameters there is a noticeable effect on the amplitude and position of the CRR features. For $\aEM$, the modifications are more pronounced, which in part is due to a near degeneracy of $\me$ with modifications to the time of recombination \cite{Hart2023VFCCRR}. Nevertheless, it is clear that precision measurements of the average CMB spectrum could principally identify these modifications in comparison to the standard $\Lambda$CDM prediction, thereby shedding new light on the physical cause of the problem.

However, we would like to highlight that these novel opportunities are not limited to searches for VFCs. Also EDE models (see \EDEC) indirectly affect the shape of the CRR \cite{Hart2023VFCCRR}. In Fig.~\ref{fig:h_all}, we illustrate the effect of EDE on the CRR. Importantly, depending on the parameter choices \cite{Hart2023VFCCRR}, the redshifting between the Hydrogen and Helium recombination lines can be modified in addition to when the lines are created. This leads to principally observable features in the CRR, would a sufficient spectral sensitivity be reached. 
Finally, small-scale perturbations in the baryonic density, as possibly induced by PMFs (see \PMFC), can leave an imprint to the CRR that could be used to identify this process (see Fig.~\ref{fig: PMFs}). Physically, the modifications are simply related to the fact that the average CRR is no longer given by the CRR of the average parameters, as non-linear (non-perturbative) corrections to the recombination process remain.
Overall, all these example illustrate that future CMB spectral distortion measurements could in principle allow us to directly test the underlying recombination process. From the scientific point of view it would actually be crucial to investigate the recombination history directly, thereby eliminating one of the remaining theoretical ingredients of in CMB cosmology.
However, it is also clear that due to the presence of foregrounds the required spectral sensitivity and coverage are still quite futuristic \cite{Mayuri2015, Vince2015, Hart2020CRR}. In addition, a control of systematics and removal of foregrounds will have to be performed to unprecedented precision. These challenges will remain to be solved by generations of cosmologists to come.

\vspace{-3mm}
\section{Conclusions}
\label{sec:conclusions}
In this chapter, we illustrated the role of VFCs in the Hubble tension. Previous studies have shown that varying $\me$ could indeed offer viable solutions, although no VFC scenario could fully resolve all aspects of the tension. It is however important to note that additional studies in the non-perturbative regime (i.e., significant changes with explicit time-dependence) could further improve the consistency. In addition, so far no attempt has been made to simultaneously treat VFCs and EDE, although both could principally originate from the same scalar field. With the advent of improved cosmological data, these lines of research might become very interesting.

One obvious question remains: how can we ultimately distinguish between various solutions to the Hubble tension. As highlighted in Sect.~\ref{sec:CRR}, EDE, PMFs and VFCs can in principle also be (directly) constrained through detailed measurements of the CRR. This guaranteed $\Lambda$CDM signal may become observable in the future with advanced CMB spectrometers and would open a way to confront our understanding of the cosmological recombination process with direct observational evidence. Should the Hubble tension not be resolved in the next decades, this future probe might be one of the important avenues towards a final primordial test. 

\begin{acknowledgement}
We thank Yacine Ali-Ha\"imoud for useful comments on the manuscript.
This work was supported by the ERC Consolidator Grant {\it CMBSPEC} (No.~725456).
JC was furthermore supported by the Royal Society as a Royal Society University Research Fellow at the University of Manchester, UK (No.~URF/R/191023).
Part of this work was carried out in the stimulating and peaceful environment offered at the Aspen Center for Physics, which is supported by the National Science Foundation grant PHY-2210452.
\end{acknowledgement}


\small
\bibliographystyle{abbrv}
\bibliography{Lit}

\end{document}